\pdfoutput=1
\documentclass{article}

\usepackage[T1]{fontenc}
\usepackage{times}
\usepackage{graphicx}
\usepackage{amsmath}
\usepackage{array}
\usepackage{booktabs}
\usepackage[numbers,square,sort&compress]{natbib}
\usepackage{xurl}
\usepackage[hidelinks]{hyperref}

\newif\ifgrconkit
\IfFileExists{grcon.sty}{\grconkittrue}{\grconkitfalse}
\ifgrconkit
  \usepackage[accepted]{grcon}
  \makeatletter
  \renewcommand{\GRCON@appearing}{\textit{Proceedings of GNU Radio Conference 2026},
  Copyright 2026 by the author(s).}
  \makeatother
  \grcontitlerunning{\scriptsize gr-ntnisac: a GNU Radio NTN-ISAC Testbed}
\else
  \usepackage[margin=0.75in]{geometry}
  \newcommand{\grcontitle}[1]{\begin{center}{\Large\bfseries #1}\end{center}}
  \newcommand{\grconauthor}[2]{\begin{center}{\normalsize #1 \quad #2}\end{center}}
  \newcommand{\grconaddress}[1]{\begin{center}{\small #1}\end{center}}
  \newcommand{\grconkeywords}[1]{\begin{center}\parbox{0.9\textwidth}{\small\textbf{Keywords:} #1}\end{center}}
\fi
\setcitestyle{numbers,square,citesep={,}}
\newcommand{\code}[1]{\texttt{#1}}

\begin{document}
\sloppy

\twocolumn[
\grcontitle{gr-ntnisac: An Open GNU Radio Testbed for NTN-ISAC}

\grconauthor{Soham Dhiren Desai}{\href{mailto:soham.d.desai@gmail.com}{soham.d.desai@gmail.com}}
\grconaddress{Skylo Technologies, California, USA}

\grconkeywords{GNU Radio, 5G NR, non-terrestrial networks, integrated
sensing and communications, LEO channel emulation, passive radar, USRP,
hardware in the loop}
]

\vskip 0.3in

\begin{abstract}
A passive receiver can sense nearby objects from the echoes of a 5G
downlink.  When the transmitter is a low-Earth-orbit (LEO) satellite, one
transmitter covers a wide area, but the receiver must handle tens of kHz of
Doppler and a direct-path delay that changes within every observation
interval.  Such a receiver should be tested where the correct answer is
known before it is tried on a satellite.  This paper presents
\emph{gr-ntnisac}, an open GNU Radio testbed that applies an emulated LEO
channel to a standard 5G New Radio downlink, sends it through two cabled
USRPs, and scores the receiver against the delay and Doppler commanded for
every path.  The same receiver code runs in simulation, on recorded
captures, and on hardware.  As a first use, we measure cancellation of the
direct path, the strong satellite signal that masks weak echoes.  Depth is
34.8~dB on hardware and 37.6~dB in a matched simulation without radios, so
the radios are not the main limit.  Controlled simulation shows that the
emulator's 8-tap fractional-delay filter and a simulated rotating blade at
almost the direct path's delay each limit depth to about 40~dB; without
either, depth reaches 71.0~dB.  A diagnostic canceller that also removes
the per-subcarrier response and per-symbol phase reaches 72.7~dB on the
same captures.  On a conducted capture, a payload-aided receiver detects a
$-25$~dB moving path in 76.9\% of dwells.
\end{abstract}

\section{Introduction}
\label{sec:intro}

\begin{figure}[!b]
  \centering
  \includegraphics[width=\columnwidth]{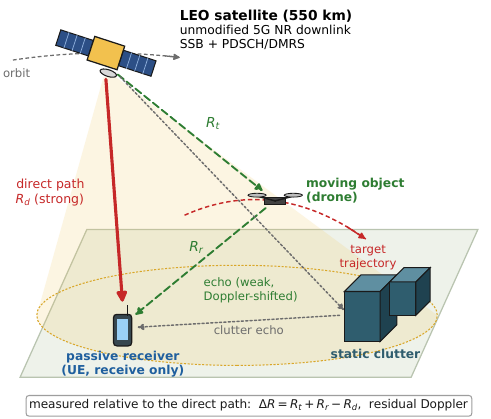}
  \caption{NTN-ISAC scenario emulated by \emph{gr-ntnisac}.}
  \label{fig:system}
\end{figure}

\begin{figure*}[t]
  \centering
  \includegraphics[width=\textwidth]{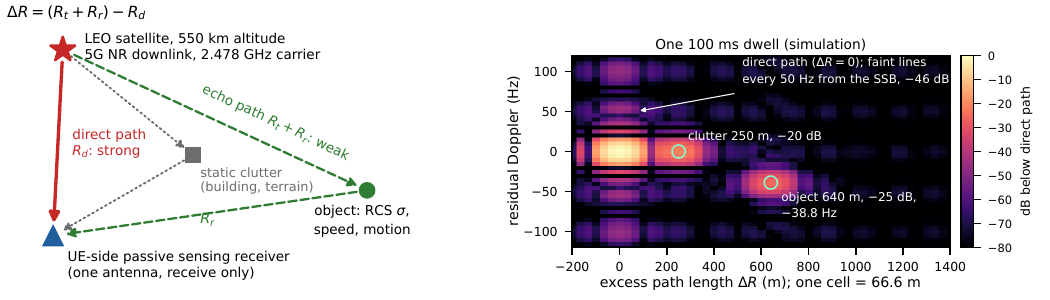}
  \caption{Bistatic geometry (left) and the delay--Doppler map of one simulated dwell (right).}
  \label{fig:geometry}
\end{figure*}

A 5G base station transmits a wideband downlink of known reference signals
and payload data.  Objects near a receiver reflect part of it, so a
receiver that detects the echoes can sense its surroundings without any
change to the transmitter: a form of integrated sensing and communications
(ISAC) \citep{liu2022isac}.  3GPP has written service requirements for 5G
sensing \citep{3gpp22137}, and the Next G Alliance treats ISAC as a
foundational capability of 6G \citep{nga2025isac}.

In a non-terrestrial network (NTN), the transmitter is on a LEO satellite
that illuminates a very large area
\citep{you2024leoisac,blazquez2024capabilities}.  The receiver must handle
tens of kHz of Doppler, hundreds of Hz/s of Doppler rate, and a direct-path
delay that changes within every observation interval
\citep{blazquez2024capabilities}.  Simulation tests the algorithms but omits
converter filters, oscillator phase noise, and clock offsets.  A satellite
trial includes them, but gives no repeatable path-level truth and no matched
target-absent control.  Between the two, a lab setup is needed that runs the
receiver on real radios while the delay and Doppler of every path are known.

Terrestrial 5G passive radar has been demonstrated with the synchronization
signal block (SSB) alone \citep{abratkiewicz2023ssb}, with the full
downlink \citep{wypich2026experimental}, and on deployed networks
\citep{samczynski2022passive}, and an open real-time OFDM-ISAC platform
exists for terrestrial links \citep{zhou2026openisac}.  For LEO
illuminators, the published work known to us is analytical or simulated;
we are not aware of an open hardware-in-the-loop testbed.

\emph{gr-ntnisac} provides one.  It applies the LEO channel to a standard
5G New Radio (NR) downlink in software, transmits the result through real
radios, and compares the receiver's output with the path trajectories
written in a scene file.  Real converters, oscillators, and host streaming
are in the loop, and every path's commanded delay and Doppler remain known
on the sample clock.  The same receiver code runs in simulation, on
recorded captures, and on hardware, so a result that changes between steps
points to the effect that step adds.  GNU Radio suits this because stream
processing, message passing, hardware access, and replay share one
flowgraph \citep{blossom2004gnuradio}.

Figure~\ref{fig:system} shows the scenario.  The receiver gets a strong
\emph{direct path} and much weaker \emph{echoes} from moving objects and
static clutter.  It measures each echo relative to the direct path: its
\emph{excess path length} $\Delta R=(R_t+R_r)-R_d$ from the extra delay,
and its \emph{residual Doppler} from the rate at which $\Delta R$ changes.
Each delay--Doppler map is computed from one observation interval, a
\emph{dwell}, here 100~ms.  Figure~\ref{fig:geometry} shows a bench
simulation with a static scatterer at 250~m and a $-25$~dB drone whose
excess path grows at 4.7~m/s; both appear where commanded, the drone at
640~m and $-38.8$~Hz.

The contributions are:
\begin{itemize}
\item \emph{LEO channel:} a real-time C++ channel in which range is the only
kinematic input, so Doppler, Doppler rate, and delay walk stay consistent;
a truth block on the sample clock; and per-symbol retiming that removes the
direct path's delay walk (\S\ref{sec:onerule}, \S\ref{sec:retime}).
\item \emph{Sensing scenes:} 3GPP models terrestrial sensing channels
(TR~38.901, Clause~7.9) and NTN propagation (TR~38.811) separately
\citep{3gpp38811,3gpp38901}; a scene file places sensing objects and 3GPP
multipath under an analytic or SGP4 orbit with commanded truth
(\S\ref{sec:scene}).
\item \emph{Payload-aided references:} known references are a small part of
the downlink \citep{zhang2021overview,wypich2026experimental}; the receiver
compares reference-only, payload-aided, and genie references on the same
dwell (\S\ref{sec:refs}, \S\ref{sec:payload}).
\item \emph{Hardware in the loop:} two radios add frequency, timing, and
sample-clock offsets \citep{zhou2026openisac}; every capture is recorded
for replay and compared with matched and controlled simulations
(\S\ref{sec:gates}, \S\ref{sec:usecase}).
\end{itemize}

The channel and objects are emulated and the radios are cabled, so the
testbed does not test real satellite reception or distant targets; it tests
the receiver on real hardware against known truth.  Installation and
operating instructions are provided with the public code
\citep{ntnisac2026software}.

\section{Architecture}
\label{sec:arch}

\begin{figure*}[t]
  \centering
  \includegraphics[width=\textwidth]{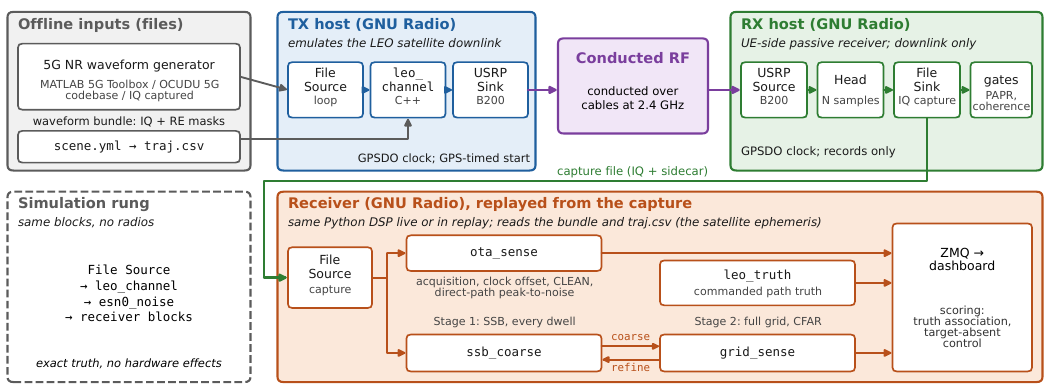}
  \caption{End-to-end testbed: emulated LEO channel, two cabled USRPs, and a replay receiver.}
  \label{fig:testbed}
\end{figure*}

Figure~\ref{fig:testbed} shows the system.  Three design decisions shape it.

\emph{The waveform is generated offline.}  The released waveform is one
20~ms NR frame (20 slots, 153,600 samples) written with the MATLAB 5G
Toolbox \citep{matlab5g} following TS~38.211 \citep{3gpp38211}, with the
reference values and occupancy masks the receiver needs.  The transmitter
repeats this frame.  Slot~0 carries the SSB and no data; slots 1--19 carry
QPSK data on all 25 resource blocks.  Because the occupancy changes once
per frame, every strong path has weak Doppler sidelobes at multiples of
50~Hz (about 46~dB below the direct path in Figure~\ref{fig:geometry}).  A
live downlink with a 20~ms SSB period has the same frame-rate structure.
The repeated payload is a simplification that the payload-aided receiver
does not exploit: it re-creates the data in every dwell from its own
decisions.  The receiver contains no NR sequence generation, so another
standards-compliant source or recorded IQ can replace the generator.

\emph{The satellite channel is applied digitally on the transmit side},
before the digital-to-analog converter (DAC).  The radios then add their
converters, filters, and clocks without changing the commanded
trajectories used for scoring.

\emph{Displays are separate processes.}  Detections and truth leave the
flowgraph over ZeroMQ, so a dashboard failure cannot stop a capture.

Table~\ref{tab:blocks} lists the blocks.  Every Python block is a thin
wrapper around the \code{py/ntn\_isac} library, which the offline tests
call directly, so tests and flowgraphs use the same receiver functions.
The \code{ota} prefix is historical: those blocks were first written for
two free-running radios.

\begin{table}[t]
\caption{Blocks of \code{gr-ntnisac}}
\label{tab:blocks}
\centering
\footnotesize
\setlength{\tabcolsep}{3pt}
\begin{tabular}{@{}l>{\raggedright\arraybackslash}p{0.62\columnwidth}@{}}
\toprule
Block & Function \\ \midrule
\code{leo\_channel} (C++) & Applies each path's range history and gain \\
\code{leo\_truth} & Each scored path's commanded excess range and Doppler,
time-stamped on the sample clock \\
\code{esn0\_noise} & Noise referenced to energy per occupied RE \\
\code{ssb\_coarse} & Stage~1: SSB matched filter, delay gate, Kalman
track \\
\code{grid\_sense} & Stage~2: full-grid delay--Doppler map, CLEAN, CFAR,
clustering \\
\code{ota\_sense} & Acquisition, clock tracking, direct-path CLEAN,
micro-Doppler \\
\code{link\_budget} & Echo-to-direct ratio for a cross section and range \\
\code{msg\_print} & Text sink for any message port \\
\bottomrule
\end{tabular}
\end{table}

\section{Channel and Scenes}
\label{sec:channel}

\subsection{Range Is the Only Input}
\label{sec:onerule}

Each path $p$ is described only by its range $R_p(t)$ and gain over time,
read from a trajectory table.  The channel computes
\begin{equation}
y(t)=\sum_{p} a_p(t)\,x\!\left(t-\tau_p(t)\right)
e^{-j2\pi f_c\tau_p(t)},
\label{eq:channel}
\end{equation}
where $x$ is the baseband waveform, $y$ the output sent to the DAC, $a_p$
the path amplitude, $\tau_p=R_p/c$ its delay, and $f_c$ the carrier.  As
the range changes, the carrier phase rotates, which is the Doppler shift;
Doppler is not a parameter anywhere in the code.  A channel with separate
delay and Doppler inputs could describe an object whose range and Doppler
disagree, and a frequency-shift-only model omits the \emph{delay walk}, the
change of delay during a dwell.  For the pass in Table~\ref{tab:pass}, the
direct path's delay changes by up to 2.2~$\mu$s (17 samples at
7.68~MS/s) within one 100~ms dwell.  One delay bin of the 5~MHz waveform
(4.5~MHz occupied) is 222~ns, or 66.6~m of path, so the direct path
migrates across about 10 bins per dwell.

\begin{table}[t]
\caption{Sample SGP4 Pass}
\label{tab:pass}
\centering
\footnotesize
\begin{tabular}{@{}>{\raggedright\arraybackslash}p{0.45\columnwidth}>{\raggedright\arraybackslash}p{0.49\columnwidth}@{}}
\toprule
Quantity & Value \\ \midrule
Orbit / pass & 550~km circular, 600~s, 67.5$^\circ$ peak elevation \\
Slant range & 594 to 2170~km \\
Range rate / Doppler & $\pm6.636$~km/s / $\pm54.9$~kHz \\
Doppler rate & up to 678~Hz/s \\
Delay walk, 7.68~MS/s & 170 samples/s \\
\bottomrule
\end{tabular}
\end{table}

The direct path comes from an analytic circular orbit in Python or from
MATLAB SGP4 propagation of a two-line element set (TLE); both write the
same trajectory format.  Table~\ref{tab:pass} is an SGP4 pass of a
synthetic 550~km TLE over Mountain View, California.

\subsection{Implementation}
\label{sec:cpp}

The C++ block evaluates Equation~\eqref{eq:channel} for every sample, with
the delay from a cubic spline through the trajectory and the carrier phase
reduced modulo one cycle in double precision.  An interpolation filter
supplies each fractional-sample delay.  Its frequency response is not flat
and varies with fractional delay, which puts sidebands on the direct path;
longer filters reduce them at a cost in throughput.  The 5~MHz waveform is
sampled at 7.68~MS/s (512-point FFT, 15~kHz subcarrier spacing).  With the
8-tap windowed-sinc filter used on hardware, the block measured 20.4~MS/s
on three paths, 2.66 times real time.  This filter's response reappears in
the hardware results.

\subsection{Scenes}
\label{sec:scene}

A scene is a YAML file listing the orbit and every path
(Figure~\ref{fig:scene}); \code{tools/build\_scene.py} compiles it into a
trajectory.  Path types are \code{direct}, \code{static}, \code{linear},
\code{rotating} (a point on a blade, for micro-Doppler), \code{pendulum},
\code{custom} (an expression in $t$), and \code{ntn\_tdl}.  Any path can be
limited to a visibility window, giving target-absent epochs inside one
capture, and any field can be overridden on the command line, so a sweep
is a shell loop.

\begin{figure}[t]
\footnotesize
\begin{verbatim}
name: bench_both
carrier_hz: 2.478e9
orbit: {source: circular, alt_km: 550,
        max_elev_deg: 67.5}  # or SGP4
paths:
  - {type: direct}
  - {type: static, excess_m: 250,
     gain_db: -20}
  - {type: linear, excess_m: 640,
     excess_rate_mps: 4.7, t_ref_s: 500,
     gain_db: -25}
  - {type: rotating, excess_m: 0.4,
     radius_m: 0.25, rot_hz: 7,
     gain_db: -30}
noise: {esn0_db: 45}
\end{verbatim}
\caption{Scene file \code{scenes/bench\_both.yml}, condensed.}
\label{fig:scene}
\end{figure}

\emph{Object size.}  An echo's gain is its power relative to the direct
path.  For a distant transmitter it is approximately $\sigma/(4\pi R_r^2)$
for radar cross section $\sigma$ at distance $R_r$ from the receiver,
which \code{link\_budget} computes.  For example, $-25$~dB is 400~m$^2$
at 100~m, and $-40$~dB is 12.6~m$^2$ at 100~m or 1,260~m$^2$ at 1~km.
Cross sections can come from electromagnetic simulation tools.

\emph{Geometry.}  Objects near the line of sight have little excess path:
one meter off the middle of a five-meter baseline gives 0.39~m, under 1\%
of a delay bin.  Such objects are separated by Doppler, which is why the
bench scenes include a rotating target.

\emph{3GPP multipath.}  The scene imports the TR~38.811 tapped-delay-line
profiles \citep{3gpp38811}.  A tap with a time-varying complex gain would
reintroduce a Doppler not constrained by range, so each cluster is built
from sub-rays whose range ramps at the rate that gives their Doppler.  With
several sub-rays spread in angle, as in Jakes' model
\citep{jakes1974microwave}, the cluster fades by interference, with no new
channel code.  These clusters carry no truth and are switched off when
sensing is scored.

\section{Receiver}
\label{sec:rx}

\subsection{OFDM Sensing}
\label{sec:ofdm}

Each OFDM symbol carries one complex value on each of 300 subcarriers
spaced 15~kHz apart; one subcarrier in one symbol is a \emph{resource
element} (RE).  Each symbol is preceded by a \emph{cyclic prefix} (CP), a
copy of its last portion.  At 15~kHz with normal CP and 7.68~MS/s, most
symbols have a 36-sample (4.69~$\mu$s) CP, while symbols 0 and 7 of each
14-symbol slot have 40 samples \citep[Sec.~5.3.1]{3gpp38211}.  An echo
arriving within the CP after the direct path is a circularly shifted copy
of the symbol, so after the FFT its delay is a linear phase across
subcarriers and its Doppler a linear phase across symbols.  Once the
transmitted values are divided out, two FFTs give the delay--Doppler map
\citep{sturm2011waveform}.  Bins are 66.6~m by 10~Hz for a 100~ms dwell,
the shortest CP bounds the excess path at 1405~m, and coherent use of all
374,650 occupied REs gives 55.7~dB of integration gain.

\subsection{Retiming and De-Walk}
\label{sec:retime}

The goal is to hold the direct path at zero delay and Doppler in every
symbol.  Its Doppler reaches $\pm55$~kHz, almost four subcarrier spacings,
and its delay walks up to 17 samples (10 bins) per dwell; a window placed
once per dwell would smear every path across those bins and use up half
the CP.  An NTN user equipment pre-compensates from the broadcast ephemeris
\citep{3gpp38331,3gpp38300}, and radar calls this range migration
\citep{perry1999keystone}.  Here the trajectory file is the ephemeris.  The
receiver removes the direct path's carrier phase, which takes out its
Doppler and Doppler rate; places each symbol's FFT window at that symbol's
arrival sample, found by two fixed-point iterations because delay must be
evaluated at arrival time; and removes the remaining fractional sample as
a phase ramp after the FFT.  Echoes walk only 0.5~m per dwell relative to
the direct path at 5~m/s, so they need no further correction.

\subsection{Sensing References}
\label{sec:refs}

The receiver must know the transmitted values to divide them out
\citep{sturm2011waveform,berger2010passive}.  \emph{Reference-only} uses
the known reference signals, 8.1\% of the occupied REs.
\emph{Payload-aided} also re-creates the data REs from its own channel
estimate, equalizer, and hard decisions.  \emph{Genie} uses the transmitted
grid; it exists only in emulation and bounds the payload-aided result.
All three share one code path, with untrusted REs masked and the map
normalized by the number of REs used.  Payload reconstruction is
established in passive radar \citep{poullin2005passive,palmer2013dvbt} and
is not claimed as new.

\subsection{Cancellation and Detection}
\label{sec:detect}

The direct path is at least 20~dB stronger than every modeled echo, so it
is removed first.  From the reference grid and occupancy mask, the receiver
builds the ideal point-spread function of a hypothesized path, finds and
refines the direct-path peak, fits its amplitude by least squares, and
subtracts it: H\"ogbom's CLEAN \citep{hogbom1974aperture,kulpa2008clean},
using the actual NR symbol times and a bounded search so that a sideband or
target is not removed first.  Detection then applies an ephemeris-derived
clutter mask, cell-averaging CFAR, density clustering
\citep{ester1996dbscan}, and rejection of the 50~Hz frame-rate sidelobes.
The CFAR is set for about one noise-only candidate per dwell
\citep{richards2014fundamentals}; deterministic residue can violate this.
The LEO-specific parts are per-symbol retiming with fractional de-walk, a
CLEAN built from the exact occupancy and symbol times, and one trajectory
file serving as both ephemeris and truth.

\subsection{Two-Stage Processing}

Following \citet{khawaja2026twostage}, a low-cost stage runs on every dwell
and a full stage on demand.  \code{ssb\_coarse} matched-filters each SSB
and publishes a delay gate; \code{grid\_sense} runs the full-grid chain
inside it.  Messages carry sample-clock time stamps, so a late result is
applied to the right dwell.  GNU Radio runs each block in its own thread;
in Python, Stage~1 takes 82~ms per 100~ms dwell and keeps pace, while
Stage~2 takes about 480~ms and live runs at most once every five dwells.
All results here are from replay, where Stage~2 runs on every dwell.

\section{Method}
\label{sec:howto}

\begin{figure}[t]
  \centering
  \includegraphics[width=0.78\columnwidth]{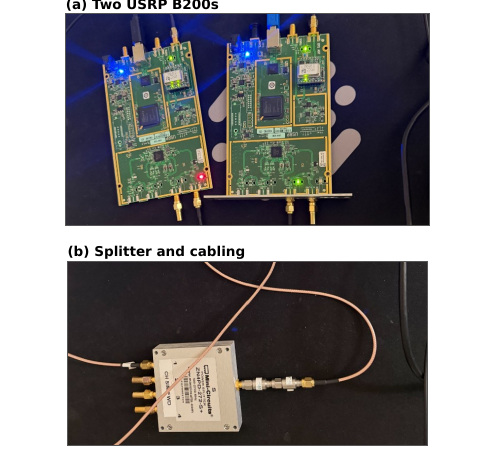}
  \caption{Conducted hardware setup.}
  \label{fig:hardware}
\end{figure}

Experiments climb three rungs with the waveform, scene, and receiver held
fixed, so each step isolates the effects it adds.  The \emph{bench} rung
runs the whole chain on one host.  \emph{Loopback} connects one radio's
ports through at least 40~dB of attenuation, checks sustained streaming,
and measures the fixed hardware delay.  The \emph{conducted} rung uses two
hosts, two USRP B200s, a splitter, and fixed attenuation
(Figure~\ref{fig:hardware}); each radio has its own GPS-disciplined
oscillator and the capture starts on a common GPS epoch.  Gain keeps the
direct path 15--20~dB below full scale.  The carrier is 2.478~GHz, not an
NTN allocation, and every hardware run is conducted over cable.

\emph{Capture, then replay.}
\label{sec:gates}
The Python sensing chain can run slower than real time and back-pressure
the USRP source, and dropped samples can corrupt delay and Doppler while
level and spectrum look normal.  The receive host therefore records with a
minimal flowgraph and sensing runs on replay.  Receive gain is fixed.  Each
capture must pass a peak-to-average gate, within 1.5~dB of the waveform's
own value, which catches a starved transmitter (ratio rises) or clipping
(ratio falls), and a frame-to-frame coherence gate that detects a sample
splice.

\section{Hardware Results}
\label{sec:usecase}

\subsection{Direct-Path Cancellation}
\label{sec:cancel}

The conducted captures used the full-allocation 5~MHz waveform.  The
direct-path peak-to-noise ratio was $82.49\pm0.70$~dB over 28 dwells
($\pm$ is one standard deviation), about 33~dB of $E_s/N_0$ per occupied
RE after the 55.7~dB integration gain and 6.0~dB window loss.  This high
level isolates hardware effects; it does not test sensitivity at an NTN
operating point, which is a natural next step on the same bench.

\begin{figure*}[t]
  \centering
  \includegraphics[width=\textwidth]{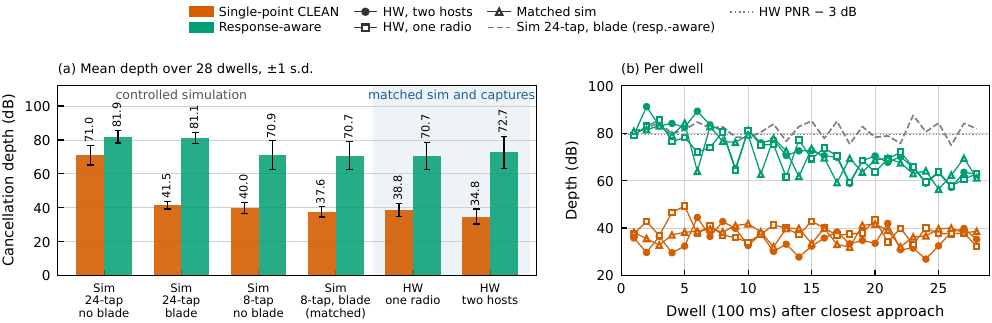}
  \caption{Direct-path cancellation depth: (a) by configuration, (b) per dwell. Response-aware: CPE (common phase error) correction, CLEAN, and a per-subcarrier term.}
  \label{fig:ladder}
\end{figure*}

Cancellation depth is the ratio, in dB, of the largest map value within
$\pm1.5$ delay bins and $\pm15$~Hz of the direct path before and after
cancellation.  With the \code{ota\_sense} configuration (Blackman--Harris
window, two CLEAN components), single-point depth was $38.71\pm3.84$~dB
with one radio in loopback and $34.75\pm4.46$~dB with two.
Figure~\ref{fig:ladder} comes from an offline MATLAB replay of the same
receiver, which reproduces these depths within 0.05~dB.  A \emph{matched
simulation} repeats the two-host experiment without radios: same scene,
waveform, and 8-tap filter, the measured clock offset of
$-1.28\times10^{-9}$ between the two oscillators, and noise set to the
captured peak-to-noise ratio.  It reaches $37.6\pm3.1$~dB, between the two
hardware values, so the radios are not the main cause.

Controlled simulation separates the causes (Figure~\ref{fig:ladder}a).
With a 24-tap interpolator and the blade switched off, single-point CLEAN
reaches 71.0~dB.  Either the 8-tap filter (40.0~dB) or the blade alone
(41.5~dB) limits depth to about 40~dB.  The blade sits at 0.4~m of excess
path, inside the direct path's delay bin, so part of the ``residue'' is a
real echo; the filter's response, which varies with fractional delay, is
the other part.

Two additions diagnose these on the captures.  The first removes the
direct path's phase in each symbol, the common phase error (CPE) that
oscillator phase noise puts on every subcarrier.  The second, after CLEAN,
fits and subtracts on each subcarrier a term at the direct path's Doppler,
approximating the combined response of the filter, radios, cables, and
splitter as constant within a dwell.  With both, the \emph{response-aware
canceller} reaches $72.7\pm9.3$~dB on two hosts, $70.7\pm7.9$~dB with one
radio, and $70.7\pm8.2$~dB in the matched simulation.  Its depth falls from
about 80 to 60~dB over the 28 dwells (Figure~\ref{fig:ladder}b) on
hardware and in the matched simulation, but not with the 24-tap filter, so
the large spreads reflect a trend more than scatter.  The captures start at
closest approach, where the range rate is zero; by the last dwell it is
250~m/s, and the fractional delay moves 0.6~sample within a dwell, while
the per-subcarrier term assumes a constant response.  Because the
common-phase step also removes slow-time modulation at the direct path's
delay, the blade included, this canceller locates the residue; it is not a
receiver for objects at that range, and since it fits and scores the same
dwell, its depth is a diagnostic upper bound.  In the same replay it cut
false detections in a target-absent capture from 4.20 to 1.28 per dwell,
close to the nominal one.

\subsection{Payload-Aided Example}
\label{sec:payload}

References alone use 8.1\% of the occupied REs, giving up 10.9~dB of
coherent energy, and the DMRS on every other subcarrier adds grating
sidelobes in delay.  Comparing references is one flag (\code{--variant
reference-only}, \code{payload-aided}, or \code{genie}).  On the bench rung
with the SGP4 pass and $E_s/N_0=10$~dB, a moving $-25$~dB object stood
9.6~dB above its CFAR threshold with references only, 13.0~dB
payload-aided, and 13.1~dB with the genie.  The payload-aided map gains the
full 10.9~dB in peak-to-noise ratio (26.5 to 37.4~dB), but the margin rises
only 3.4~dB: the direct path is not cancelled in this example, so once
noise is low the CFAR threshold is set by deterministic sidelobes.  At
40~dB $E_s/N_0$ the margins saturate at 10.6 and 13.5~dB.  Decision errors
cost little: the uncoded symbol error rate is 0.4\% at 10~dB.

On hardware, a 30~s conducted capture carried a ladder of four moving paths
at $-25$, $-30$, $-35$, and $-40$~dB and a static path at 1050~m.
Replayed through the GNU Radio blocks, the payload-aided receiver
associated the $-25$~dB path in 230 of 299 dwells (76.9\%) within one delay
bin and 10~Hz, with errors of $-4.2\pm4.0$~m and $0.9\pm0.5$~Hz.  The
weaker paths were not detected, the static path was suppressed as clutter,
and a matched target-absent capture gave no association in 299 dwells.
False detections remain high: 1,442 in the ladder capture (4.8 per dwell),
1,163 of them at multiples of 50~Hz, and 894 (3.0 per dwell) in the
target-absent capture, 688 on the same comb.  They come from direct-path
residue and the frame structure, not from targets.

\section{Future Work}
\label{sec:limits}

The testbed models only a common reference error, in which carrier and
sample clock move together; a transparent satellite payload adds a carrier
offset with no matching delay drift.  An impairment block for carrier
offset, phase noise, and amplifier nonlinearity, and a clock tracker in
\code{grid\_sense}, come next.  On the receiver side: the response-aware
canceller in the live blocks, stronger rejection of the 50~Hz comb, and
decoder-aided references; on the emulator side, a longer filter.
Extending the channel over several synchronized USRPs would allow
multistatic and multi-antenna receivers, and emulating several satellites
with handover would let a receiver track an object across passes.  A
non-repeating payload or live gNB source, radiated tests, 20~MHz hardware
runs, and a broader payload-aided evaluation are left to a continuation
study.

\section{Conclusion}

\emph{gr-ntnisac} lets an experimenter test an NTN sensing receiver in
simulation and on real radios before a satellite trial, scored against
commanded path truth with the same code throughout.  One channel rule,
range as the only kinematic input, covers moving and rotating targets,
3GPP multipath, and common reference error.  On conducted hardware,
cancellation reached 34.8~dB, close to the 37.6~dB of a matched simulation,
and controlled simulation traced the limit to the emulator's 8-tap filter
and an unresolved blade echo.  A payload-aided receiver detected a
$-25$~dB path in 76.9\% of conducted dwells, while standing false
detections from residue and the 50~Hz comb remain the main receiver
limitation.  This is the purpose of the tool: to make differences between
simulation and hardware visible, repeatable, and attributable.

\begingroup
\footnotesize
\bibliographystyle{unsrtnat}
\bibliography{references}
\endgroup

\end{document}